\documentclass{ws-procs9x6}
\usepackage{amsmath}
\usepackage{graphicx}

\newcommand{\fslash}[1]{\ooalign{\hfil/\hfil\crcr$#1$}}
\newcommand{\fsl}[1]{\fslash{#1}}
\newcommand{\sv}{\vec\sigma}

\def\aJLone<#1,#2>{#1}
\def\aJLtwo<#1,#2,#3>{#2}
\def\aJLyear<#1,#2,#3,#4>{#3}
\def\aJLpage<#1,#2,#3,#4>{#4}
\def\JL#1{{\it \aJLone<#1>}\ {\bf \aJLtwo<#1>}, \aJLpage<#1> (\aJLyear<#1>)}
\def\aJpage<#1,#2,#3>{#3}
\def\andvol#1{{\bf \aJLone<#1>}, \aJpage<#1> (\aJLtwo<#1>)}

\def\PR#1{{\it Phys.\ Rev.}\ \andvol{#1}}
\def\PRL#1{{\it Phys.\ Rev.\ Lett.}\ \andvol{#1}}
\def\PL#1{{\it Phys.\ Lett.}\ \andvol{#1}}
\def\NP#1{{\it Nucl.\ Phys.}\ \andvol{#1}}

\begin{document}

\title{Why do we need instantons in strangeness hadron physics?}

\author{Makoto Oka}

\address{Department of Physics, H-27, Tokyo Institute of Technology \\ 
        Meguro, Tokyo 152-8551, Japan\\
		E-mail: oka@th.phys.titech.ac.jp}

\maketitle

\abstracts{Roles of instantons in strangeness hadron physics are discussed.
After introducing general features of instantons, hadron spectroscopy under the influence of
instanton-quark effective interactions is discussed. Emphases are on the H dibaryon, spin-orbit
splittings of P-wave baryons and baryon-baryon interactions and spectrum of the pentaquarks.}

\section{Introduction}
Strangeness nuclear physics has reached a new era, which requires elevation of
phenomenological analyses of hypernuclear phenomena to a deeper understanding
of the fundamental interactions from QCD viewpoint.  Recent coordinated efforts in
experimental and theoretical researches revealed several interesting features of
hyperon-nucleon and hyperon-hyperon interactions.  For instance, it is established by now 
that the spin-orbit interaction of $\Lambda$ in nuclear medium is very weak, suggesting
cancellation of two types of spin-orbit interactions, spin-flavor symmetric and antisymmetric
ones. Discovery of pentaquarks triggers revisiting inter-quark correlations, such as diquarks
and triquarks.  It is urgent to establish foundations of such possibilities directly from QCD.

In this talk, I would like to review a new type of quark-quark interaction mediated by
instantons in the QCD vacuum. The instanton idea is not new, but its applications to
hadron spectroscopy and to hadronic interactions are being developed in these years.
I concentrate on the instanton induced interactions among quarks, and 
stress that the strangeness plays an important role as the third flavor
for the quark dynamics under the instanton induced interactions.

In sect.\ 2,  instanton and instanton induced interaction are introduced 
in the context of hadron physics.
In sect.\ 3, we discuss consequences of the instanton induced interaction in
the hadron spectroscopy and hadronic interactions. 
In sect.\ 4, we consider effects of instantons on the spectrum of 
recently discovered pentaquark baryon $\Theta^+$ and its siblings.
A conclusion is given in sect.\ 5.

\section{Instanton}

Instanton is a localized solution of the gluon equation of motion of QCD
in the 4-dimensional Euclidean space,
$D_{\mu} F_{\mu\nu} = 0$.\cite{BPST}
The solution also satisfies the self-duality (or anti-self-duality) relation
\begin{eqnarray}
F^a_{\mu\nu} =\pm \tilde F^a_{\mu\nu} \equiv \pm \frac{1}{2} \epsilon_{\mu\nu\sigma\rho} F^a_{\sigma\rho},
\end{eqnarray}
which guarantees that it satisfies the equation of motion because of the Bianchi identity,
$D_{\mu} \tilde F_{\mu\nu} = 0$. 
An important feature of the instanton solution is its nontrivial topology,
which is seen from non-zero topological charge (or winding number) defined by
\begin{eqnarray}
\nu &=& \frac{g^2}{32\pi^2} \int F_{\mu\nu}^a \tilde F_{\mu\nu}^a \, d^4x 
\end{eqnarray}
This $\nu$ takes an integer for localized $F_{\mu\nu}$ and represents the winding number
associated with the homotopy group, $\pi_3(SU(2))=Z$.
The above relations leads to
\begin{eqnarray}
S_E &=& \frac{1}{4} \int F_{\mu\nu}^aF_{\mu\nu}^a \, d^4x = \frac{8\pi^2}{g^2} |\nu| .
\end{eqnarray}

The instanton (anti-instanton) is a solution with $\nu=1$ ($\nu=-1$) with a finite action density
localized in the 4-dimensional space-time.
One of the roles of the instanton in QCD is that it mediates
quantum tunneling connecting two QCD vacua with different topologies. 
Indeed, the tunneling amplitude in the semi-classical limit is given by the instanton action,
$T\sim  e^{-S_E} \sim \exp (- \frac{8\pi^2}{g^2} |\nu| ) $.

When quarks are introduced in QCD, there appears a new role of the instanton.
The instanton acquires localized zero modes of light quarks, that is,
$\lambda=0$ solutions of the eigenvalue equation,\cite{tH}
\begin{eqnarray}
i \fsl{D} q(x) &=& \lambda q(x) .
\end{eqnarray}
It is easily seen that the right-chiral (R) and left-chiral (L) modes are paired with
the opposite sign of $\lambda$ for $\lambda\ne 0$, 
\begin{eqnarray}
i \fsl{D}  \gamma^5 q(x) &=& -i  \gamma^5 \fsl{D} q(x) = - \lambda  \gamma^5 q(x) ,
\end{eqnarray}
while for $\lambda=0$, there exists an isolated mode with a definite chirality,
\begin{eqnarray}
q_{\lambda=0} &=& \pm \gamma^5 q_{0} \quad\hbox{for $\nu=\pm 1$.}
\end{eqnarray}
One sees that the coupling of the quark zero mode 
with the instanton breaks chiral symmetry,

Chiral symmetry of the QCD Lagrangian with $N_f$ massless quarks is
$U(N_f)_L\times U(N_f)_R$.
By breaking this symmetry down to $U_f(N_f)$, we expect to have $N_f^2$ (= 9 for $N_f=3$)
Nambu-Goldstone bosons.  
In reality, one of the nine NG bosons, $\eta'$, is too heavy so that
the $U_A(1)$ symmetry must be broken.  It is realized by axial anomaly given by
\begin{eqnarray}
\partial_{\mu} J_A^{\mu 0} = 2i m_q \bar q\gamma^5q +\frac{\alpha_s}{2\pi} N_f 
{\rm Tr} [F_{\mu\nu} \tilde F^{\mu\nu} ] .
\end{eqnarray}
The anomaly term is proportional to the topological charge density of the instanton,
which indicates that the QCD vacuum with instantons can describe the anomaly effect
observed in the meson spectrum.

This role of the instanton was pointed out by 't Hooft,\cite{tH} who derived 
an effective interaction induced by the coupling of an instanton and light quarks
via their zero modes.
The strength of the interaction depends on the instanton density, 
which can be estimated from the gluon condensate of the QCD vacuum 
to be about 1 instanton/fm$^4$.
The effective interaction,  called instanton induced interaction (III), can be expressed as
\begin{eqnarray}
 L &=& \int d\rho\, n(\rho) \left[ \prod_i\left(m_i\rho - \frac{3\pi^2\rho^3}{4} \bar q^i_R q^i_L\right)
+ \hbox{(tensor terms)} \right] + \hbox{ (h.c.)}
\end{eqnarray}
where $\rho$ is the size of the instanton and $n(\rho)$ is the instanton density.

\begin{figure}[htp]
\begin{center}
\includegraphics[width=10cm]{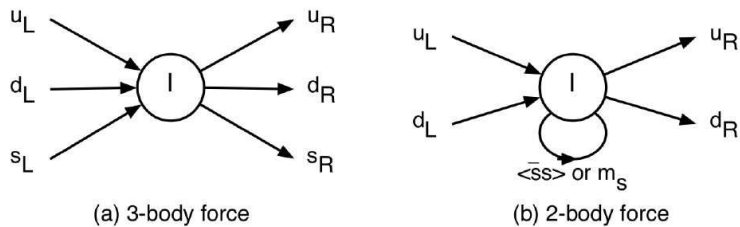}
\end{center}
\caption{Instanton induced interactions: (a) Three-body term, and (b) Two-body term.}
\label{fig:III}
\end{figure}

When the effective interaction is applied to the system of quarks, it can 
be expressed  conveniently as inter-quark potentials.  For $N_f=3$, we find\cite{OT89}
\begin{eqnarray}
V_{\rm III}^{(3)} &=& V^{(3)} \sum_{(ijk)} {\mathbb A}^f \left[ 1-\frac{1}{7} 
(\sv_i\cdot\sv_j + \sv_j\cdot\sv_k + \sv_k\cdot\sv_i) \right] \, \delta(\vec r_{ij}) \delta(\vec r_{jk}) 
\nonumber\\
V_{\rm III}^{(2)} &=& V^{(2)} \sum_{(ij)} {\mathbb A}^f \left[ 1-\frac{1}{5} 
\sv_i\cdot\sv_j \right] \, \delta(\vec r_{ij})  
\end{eqnarray} 
where $V_{\rm III}^{(3)}$ is a three-body force (6-quark vertex) and is repulsive, 
while $V_{\rm III}^{(2)}$ is a two-body term, that is mostly attractive. (Fig.~1)
The latter is generated from the three-body term by contracting a pair of quark and antiquark into
quark condensate or in the quark mass term.
In the above equations, ${\mathbb A}^f$ denotes antisymmetrization of flavor indices of the $(ijk)$ or $(ij)$ quarks.
Then the interactions are nonzero only for flavor antisymmetric states of quarks, for instance, for $[ud] (I=0)$, 
and for $[uds]$ (flavor singlet). This flavor dependence plays an important role when they are applied to hadron
spectroscopy.

Another important feature of III is that it mixes quark flavors in $\bar q q$ systems.  
It is easily seen that the III converts $\bar u u$ into
$\bar d d$ or $\bar s s$.
This provides an explicit mechanism to make the flavor singlet meson $\eta_1$ heavier 
than flavor octet mesons, $\pi$, $K$ and $\eta_8$, and thus explains the $U_A(1)$ symmetry breaking. 
Mixing III with appropriate flavor SU(3) breaking interactions, it was shown that 
pseudoscalar meson spectrum can be explained fairly well.

\section{Hadrons with instanton induced interaction}

Lattice QCD suggests\cite{Chu} that the QCD vacuum indeed contains instantons, 
whose density is consistent with the gluon condensate expected from QCD sum rules.
It was also shown that chiral symmetry is dynamically broken in the instanton vacuum.
Then massless quarks are transformed into constituent quarks,
which acquire mass as a function of momentum.
In recent developments, III is pointed out to play
a major role in diquark correlations in hadrons and quark (hadron) matter.
Antisymmetric $ud$ diquark with $I = 0$ (or flavor $\bar 3$), $J^{\pi}=0^+$ and color $\bar 3$ is
strongly favored by III.  Such diquark correlations may induce phase transition
of the QCD vacuum at high density into a color superconducting phase.

\medskip\noindent \underline{\bf\it Hyperfine interaction and H dibaryon}

The main role of III in the low-energy hadron spectroscopy is given by its spin dependence.
In the conventional quark model, splittings of the ground state octet and decuplet baryons are
supposed to come from color-magnetic terms of one-gluon exchange (OgE) interaction
between quarks. 
Indeed, if we choose the strength of the OgE interaction to be consistent 
with the $N-\Delta$ mass splitting (exp.\ 300 MeV), then it can explain 
the whole octet and decuplet spectra very well.
It is, however, too strong in the sense that the coupling strength, $\alpha_s$, thus determined
is larger than one, and perturbative expansion is not justified. 

Another problem of OgE is seen in the spectrum of 
double-strange dibaryon $H$ ($S = -2$, $B = 2$).\cite{Jaffe}
A deeply bound $H$-dibaryon was predicted, that is, the biding energy of more than 100 MeV.
A typical estimate of the H dibaryon mass in the constituent quark model gives
\begin{eqnarray}
	M_H &=& \sum m_q + \langle V_{cm} \rangle_H  
    = 360\times 4 + 540\times 2 - 450 \sim  2070 \hbox{ MeV,}
\end{eqnarray}
which is about 150 MeV above the $\Lambda\Lambda$ threshold ($\sim$ 2230 MeV).
Unfortunately, 20-year effort of searching the H dibaryon was not successful.

An alternative to OgE is the two-body part of III, which favors lower total spin.
In 1989, we proposed a picture in which a part of the hyperfine
interaction is shared by III from OgE.\cite{OT89}
This picture works because the ground state baryon spectrum is equally well reproduced by
III, because its spin dependence is almost identical to that of OgE.\cite{SR}
Even the required mass dependence of the hyperfine splitting is explained as
the effect of the strange quark mass on the instanton vacuum.
In the III+OgE quark model, the sum of the hyperfine interaction is fixed by
the splitting of the octet and decuplet baryons.
We introduce a parameter, $p_{III}$, that represents the portion of the hyperfine splitting
given by III.
For $p_{III}$ is zero, the $N-\Delta$ mass difference comes solely from OgE, while
III is responsible entirely for $p_{III}=1$.

The mixing strength may be determined in the quark model using the $U_A(1)$ breaking
in the pseudoscalar meson spectrum, that is, the $\eta-\eta'$ mass difference.  
A typical magnitude is about $p_{III} \sim 40$\%.
Then $\alpha_s$ is reduced by a factor 0.6, which is an attractive feature of this model.

It should be noted that the three-body term does not change the above picture, 
because it is effective only on flavor-singlet $u-d-s$ system. 
III that is consistent with the $N-\Delta$ mass splitting also explains the ground state baryon
spectrum in the same quality as OgE.
Flavor singlet baryon, $\Lambda^*$, is then the only place to see this effect in the ordinary 
meson-baryon spectrum.
Even in the $\Lambda^*$, it is highly suppressed at $L=1$ relative motion
because the interaction is strong only at short distance determined
by the size of the instanton ($\sim 0,3$ fm). 

In contrast, the three-body III plays a key role in the H dibaryon.\cite{TO91}
Note that H is a flavor singlet 6-quark system, and it contains multiple 
flavor-singlet $u-d-s$ components.
It is found that although the two-body III gives some attraction, 
the three-body term gives strong repulsion to push the H dibaryon above
the $\Lambda\Lambda$ threshold.

\medskip\noindent \underline{\bf\it Spin-orbit interactions}

Spin-orbit interactions of quarks and baryons provide an important clue to III.
It is well known that LS splittings of $N^*$ baryons are unexpectedly small.
This suggests that the LS force among quarks is weak.  However, when we 
consider NN interaction in terms of quark substructure of the nucleon,  it
requires much stronger LS force for quarks.  Recent studies revealed that
the one-body LS force on $\Lambda$ in hypernuclei is
much weaker than that for N.  This can be explained by cancellation of two
types of LS force, that is, flavor-symmetric (SLS) and antisymmetric (ALS) LS forces,
\begin{eqnarray}
V_{SLS} &=& V_{SLS}^0 \, (\vec\sigma_N+\vec\sigma_{\Lambda})\cdot \vec L,
\nonumber\\
V_{ALS} &=& V_{ALS}^0 \, (\vec\sigma_N-\vec\sigma_{\Lambda})\cdot \vec L  .
\end{eqnarray}
ALS is extremely small for the NN interaction because it necessarily breaks isospin
symmetry.  In contrast, it can be as strong as SLS in the YN and YY interactions.
Indeed, supposing $ V_{SLS}^0 \sim  V_{ALS}^0$ in the $\Lambda N$ LS potentials,
the one-body LS force on $\Lambda$ in hypernuclei is strongly suppressed due to
the cancellation of SLS and ALS.

\begin{figure}[htp]
\begin{center}
\includegraphics[height=7cm]{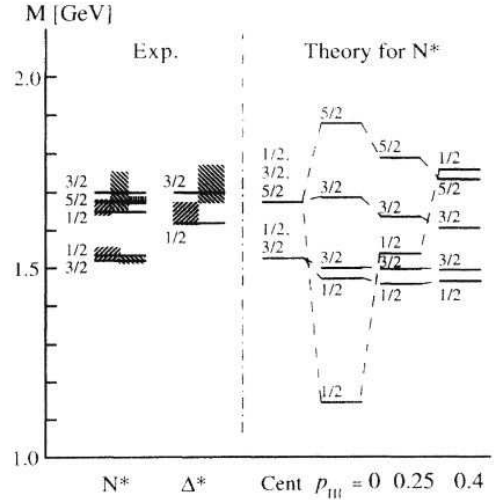}
\end{center}
\caption{P-wave $N^*$ spectrum in III compared with experiments.}
\label{fig:Nstar}
\end{figure}

Takeuchi performed the calculation of the LS force in the instanton induced interaction
and found that its flavor-dependence is different from the LS force in OgE.\cite{Takeuchi}
She pointed out that the LS splittings of the P-wave $N^*$ baryons are reduced 
significantly by mixing III (Fig.~2), while the NN and YN LS forces remain strong.
In the quark cluster model (QCM) calculation with III,  it is found that 
the antisymmetric LS force $V_{ALS}$ between $\Lambda$ and $N$ is strong, 
so that it strongly reduces the LS force for $\Lambda$.  It is also pointed out that
coupling of $N\Sigma$ plays an important role in such a calculation.

\section{Roles of Instantons in Pentaquarks}
 
Discovery of $\Theta^+$\cite{Nakano} generated a lot of interest in the
pentaquark states.  The pentaquark is a baryon which cannot be composed of three constituent
quarks.  Because $\Theta^+$ is supposed to be a $S=+1$ baryon resonance, its minimal quark
configuration is $uudd\bar s$, thus a pentaquark.

It was confirmed by several other groups later, but some negative results have been reported
from high energy $e^+e^-$ and proton induced production experiments.
Its width seems to be very small ($\sim 1$ MeV or less), which is a great mystery to be solved.
The quantum numbers other than the baryon number, strangeness and isospin ($I=0$) are not known.

A naive constituent quark model may predict a ground state with negative parity and spin 1/2,
while some quark models predict a possibly narrow negative parity state,
which decays only to KN  P-wave states.\cite{Penta04}
Dynamical calculations of the pentaquark systems in the quark model are yet to be seen, but
in general, it is not easy to bring a five-quark object down to 1.54 GeV of mass.

\begin{figure}[htp]
\begin{center}
\includegraphics[height=6cm]{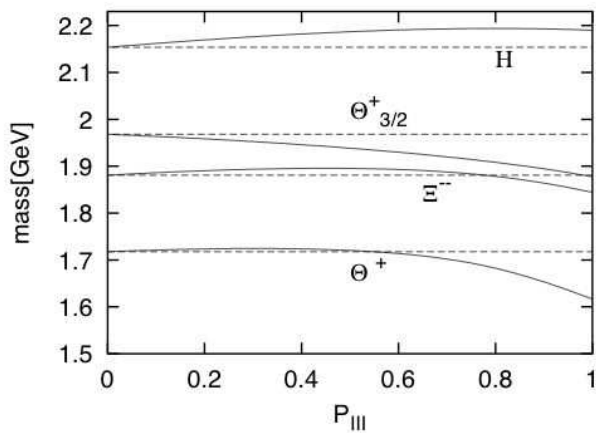}
\end{center}
\caption{Masses of pentaquark baryons and the H dibaryon as functions of $p_{III}$ calculated in the bag model. }
\label{fig:bag}
\end{figure}
Recently, an analysis of the contribution of the III for $\Theta^+$ in the MIT bag model 
was carried out by Shinozaki et al.\cite{Shinozaki}  
One important mechanism of III is that the two-body part of III is strongly attractive.
Furthermore, as its leading term is independent of spin, the attraction grows with the number
of $qq$ bonds in the system, which is ten for 5-quark systems and three for 3-quark systems.
The three-body part of III is repulsive, which amounts to 50 MeV.  This is smaller than the
repulsion in the H dibaryon, because $\Theta^+$ contains only one strange (anti)quark.
In the end, we find that the instanton gives moderate attraction ($\sim 100$ MeV)
to the negative parity pentaquark (Fig.~3).
It is also found that the spin 3/2 partner of $\Theta^+$ is strongly affected by III, while 
$S=-2$ $\Xi^{--}$, which is another possible genuine pentaquark baryon, is insensitive to III.
III also gives distinctive effects on the positive-parity pentaquark state, 
because the component favored by III is different from that by OgE.

\section{Conclusion}
The instanton represents nonperturbative effects of QCD in hadron spectroscopy and interactions.
Instanton induced interaction can be seen unambiguously in several key quantities, such as 
$\eta-\eta'$ mass difference, H dibaryon and maybe, pentaquarks.
III may also explain the spin-orbit forces in baryon spectrum and B-B interactions.
Pentaquarks have attraction from III.

\medskip
\noindent \underline{Acknowledgment}

Main part of this work was done in collaboration with Dr. Sachiko Takeuchi and Tetsuya Shinozaki.


\begin{thebibliography}{99}
\bibitem{BPST}
A.~A.~Belavin et al.,
\PL{B59,1975,85}

\bibitem{tH}
G.~'t Hooft,
\PR{D14,1976,3432},  Erratum, ibid.\andvol{18,1978,2199}.

\bibitem{OT89}
M.~Oka and S.~Takeuchi,
\PRL{63,1989,1780}

\bibitem{Chu} M.~C.~Chu et al., \PR{D49,1994,6039};
D.~Diakonov, \JL{Prog. Part. Nucl. Phys.,51,2003,173}.

\bibitem{Jaffe}
R.~L.~Jaffe,
\PRL{38,1977,195}

\bibitem{SR} E.~V.~Shuryak and J.~L.~Rosner,
\PL{B218,1989,72}

\bibitem{TO91}
S.~Takeuchi and M.~Oka,
\PRL{66,1991,1271}; 
M.~Oka and S.~Takeuchi,
\NP{A524,1991,649}

\bibitem{Takeuchi} S.~Takeuchi, \PRL{73,1994,2173}; \PR{D53,1996,6619}.

\bibitem{Nakano} T. Nakano et al., \PRL{91,2003,012002}.

\bibitem{Penta04} See papers in Proceedings of PENTAQUARK04, July 2004, SPring-8, Japan.

\bibitem{Shinozaki}
T.~Shinozaki, M.~Oka, S.~Takeuchi, hep-ph/0409103, also in Ref.~10.

\end{thebibliography}
\end{document}